# Deep leakage from gradients

Yaqiong Mu

College of Information Science and Technology, Donghua University, 201620, Shanghai China

2211826@mail.dhu.edu.cn

**Abstract.** With the development of artificial intelligence technology, Federated Learning (FL) model has been widely used in many industries for its high efficiency and confidentiality. Some researchers have explored its confidentiality and designed some algorithms to attack training data sets, but these algorithms all have their own limitations. Therefore, most people still believe that local machine learning gradient information is safe and reliable. In this paper, an algorithm based on gradient features is designed to attack the federated learning model in order to attract more attention to the security of federated learning systems.

In federated learning system, gradient contains little information compared with the original training data set, but this project intends to restore the original training image data through gradient information. Convolutional Neural Network (CNN) has excellent performance in image processing. Therefore, the federated learning model of this project is equipped with Convolutional Neural Network structure, and the model is trained by using image data sets. The algorithm calculates the virtual gradient by generating virtual image labels. Then the virtual gradient is matched with the real gradient to restore the original image.

This attack algorithm is written in Python language, uses cat and dog classification Kaggle data sets, and gradually extends from the full connection layer to the convolution layer, thus improving the universality. At present, the average squared error between the data recovered by this algorithm and the original image information is approximately 5, and the vast majority of images can be completely restored according to the gradient information given, indicating that the gradient of federated learning system is not absolutely safe and reliable.

## 1 Introduction

In modern Federated Learning (FL) systems [1-3], model updating by exchanging gradient information among multiple participants is a very common approach. The user data of each participant is always stored locally, and only the gradient information is propagated between different models. This type of algorithm does not need to establish a dedicated central node for data processing, which protects the privacy of users and the local model can be fully trained with the help of a federated learning system. For example, medical systems can share the same data model while protecting the patient's private information [4]. Therefore, it is not easy to extract the data information of local models from the gradient, which has long been believed to be able to be propagated among different models without worrying about privacy leakage, but in fact, stealing local information from the gradient is still traceable.

With the rapid development of AI technology, federation learning models are increasingly used as a fundamental technique in AI technology. Federal learning keeps the data of each participant locally, and the databases of each participant remain independent of each other during modeling, while the information interaction during joint training is encrypted to ensure the confidentiality and efficiency of the system. In addition, the federated learning system can guarantee that the training effect of the local training model is almost the same as that of the original centralized training model.

Nowadays, the development of artificial intelligence and deep learning is rapidly changing, and federated learning solves the problem that data from all parties in the previous centralized model can only be used at the central node, and ensures the privacy and confidentiality of users at each node. Federated learning is suitable for training models with large volumes of data and can be applied in a variety of contexts. Nowadays, the concept of smart cities has gained widespread attention, and federal learning models have greatly contributed to the construction of smart cities. In terms of economy and finance, it can combine data from various banks to build a model of economic fluctuation, which can better predict the future economy, etc. In terms of politics and people's livelihood, it can build a bridge between governments at all levels and the masses, realize effective information sharing between governments and the masses, build a good platform for communication between the masses and the government, and help various governments to build a good system of people's city built by the people, so that the authorities can do their work more efficiently and the masses can do their work more conveniently, etc. efficient, more convenient for the masses, etc.

The high efficiency and confidentiality of the federal learning system make it more and more widely used. However, the confidentiality of the federal model needs to be further explored, and if the data involved in the training can be restored by some means, it proves that the system still needs to be improved. With the continuous progress of artificial intelligence, the protection of Internet privacy has gradually become a hot topic of discussion. By studying the vulnerability of the system, the confidentiality of the federation learning system is gradually improved, which can also provide some new ideas for the protection of Internet privacy nowadays.

This thesis focuses on the gradient information leakage problem in convolutional neural network-based federal learning systems, and explores how to restore the original data image from the gradients containing very little information. After introducing the basic principles, the effect of Deep Leakage from Gradients (DLG) algorithm to restore the original image is studied, and certain improvements are made based on it, and finally the corresponding conclusions are drawn by comparison.

The structure of the thesis is as follows: Chapter 1 briefly introduces the research background, status and significance of this thesis, and briefly composes the content to be studied in this thesis. Chapter 2 briefly introduces the federal learning system, the structure, functions and common models of CNN, and some attack algorithms against the federal learning system. Chapter 3 mainly introduces the general principle of local information leakage, and the working principle and derivation process of DLG algorithm. Chapter 4 mainly shows the implementation of the depth gradient algorithm, analyzes the shortcomings of the algorithm, proposes improvement methods and compares them. Chapter 5 mainly integrates and summarizes the research content of this topic, presents the shortcomings and areas for improvement, and provides an outlook for the gradient attack algorithm for FL.

## 2 Related Technologies

This section introduces the basic concepts and related techniques needed to understand the reconstruction attack based on gradient features, including the introduction of the federal learning model, the convolutional neural network structure used to train the model, the related models, the role of the functions involved in the network, and some methods for gradient-based attacks.

### 2.1 Federal Learning Model

The system for federated learning [22] first utilizes an encryption-based user sample alignment technique where data owners identify the common users of each party while securing the data of their respective users in order to federate the features of these users for modeling, and the modeling training process requires federated models to secure the privacy of each local database. First, the federated model sends the public key to the local database to ensure that the local place completes the local data encryption before performing data exchange. After that, the local place transmits the data to the joint model in encrypted form. The data has been initially calculated by the local place and the gradient is calculated based on the tag value, and then the gradient is encrypted and transmitted to the joint model. The joint model combines the gradients calculated by each local model to find the total gradient value, decrypts it and sends it to each local model, so the local model can update its own model parameters according to the new gradient value and improve the optimized model. The above process is repeated until the gradient is infinitely close to the set value, which completes the training of the whole model. During the model training process, the data of each data owner is not exposed to the federated model and other local models, and the data exchange during training does not lead to data privacy threats. As a result, all parties are able to cooperate in training the model with the help of the federated learning model.

### 2.2 Convolutional Neural Networks

Convolutional Neural Network (CNN) is a deep learning model inspired by biological neural networks [23], formed by interconnecting multiple layers of neurons, where the number of input data in each layer is equal to the number of neurons in the previous layer, and each neuron can receive multiple inputs but can only output one data. This network is often applied in image processing, and the structure and role of each layer will be described next [24].

**Input Layer.**
Convolutional neural networks first need to convert image information into input data. The color of a color picture pixel consists of three attributes: red, green and blue, which are called RGB three channels, and the number of pixels in each row and column of each picture is the resolution of the picture. However, for black and white pictures, the color of the pixels is determined only by the attribute grayscale value. Assume that the value of each channel is between 0 and 511. A color photo with a resolution of 100×100 can be converted to a tensor of (100,100,3), and a black and white photo of the same size can be converted to a tensor of (100,100,1).

The main work of this layer is to perform a pre-processing of the original image, which consists of three main categories: Centering, which subtracts the average of this dimension from each dimension of the input data, so that the center of the data lies at the zero point. Normalization, which makes the standard deviation of the data to be 1, reduces the effect of different values taken by the data. PCA is used to reduce the correlation between the feature values and strives to eliminate the correlation between image bands; and whitening, which weakens the effect of the magnitude on the feature axis of the data.

**Convolutional Layer.**

| 1 | 1 | 1 | 1 | 1 |
|---|---|---|---|---|
| -1 | 0 | -3 | 0 | 1 |
| 2 | 1 | 1 | -1 | 0 |
| 0 | -1 | 1 | 2 | 1 |
| 1 | 2 | 1 | 1 | 1 |

⊗

| 1 | 0 | 0 |
|---|---|---|
| 0 | 0 | 0 |
| 0 | 0 | -1 |

=

| 0 | -2 | -1 |
|---|---|---|
| 2 | 2 | 4 |
| -1 | 0 | 0 |

**Fig. 1.** Two-dimensional convolution example

The three hyperparameters of the convolution kernel are Stride, Zero Padding and Depth. Stride is the number of frames that the data frame moves, which in Figure 2-3 is equal to 1. Zero padding protects the edge information of the image from being blurred or lost during the network training process. Depth is the number of convolution kernels, which should be the same as the number of neurons in the next layer. The number of neurons in the convolutional layer is calculated by subtracting the number of neurons from the size of the convolution plus twice the sum of the zero padding, dividing by the step size, and finally adding one to the resulting result.

Without parameter sharing, 10×64×64×5×5×3=3072000 parameters are required, and with parameter sharing, 10×5×5×3=750 parameters are required. It can be seen that parameter sharing reduces the number of features obtained by the convolutional nuclei, which leads to the loss of local features if the image size is large. An effective way to solve this problem is to set multiple convolutional kernels in each convolutional layer.

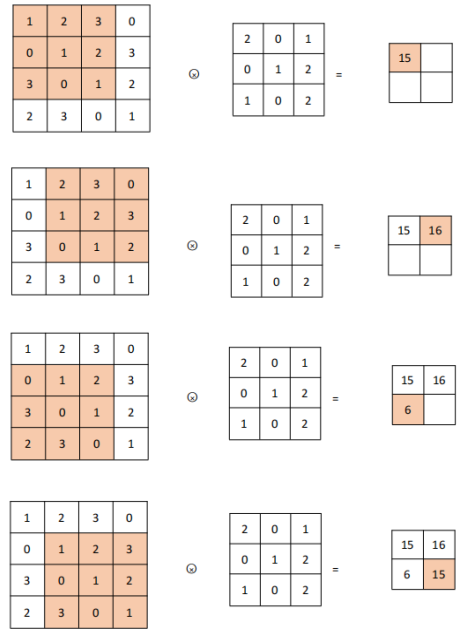

Fig. 2. Feature Mapping

**Activating Layer**

The role of this layer is, as the name suggests, both to take the output of the convolutional layer and to process it nonlinearly. Commonly used nonlinear mapping functions will be introduced in the following.

Sigmoid function

Advantages: take the value range (0, 1), simple, easy to understand.

Disadvantages: too much data may paralyze the neuron, so that the gradient information cannot be transmitted; the function output data center point does not lie at the zero point.

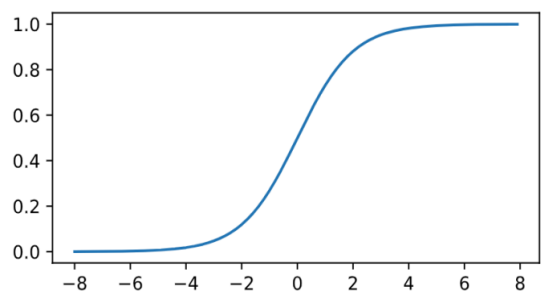

Fig. 3. Sigmoid function

**Pooling Layer**

The pooling layer, also called subsampling layer, is used for feature extraction, which reduces the number of neurons to some extent and prevents the appearance of overfitting. This layer removes redundant information and retains only key features, which can improve robustness. The pooling layer, also known as the downsampling layer, causes the features of the input information to be lost, which in turn cuts the number of parameters, making the network less computationally burdensome; while keeping the important features unchanged (cropping, stretching, scaling, etc.).

One is average pooling, which requires summing the feature points in the neighborhood and then dividing the total feature value equally among the feature points; the other is maximum pooling, which, as the name implies, excludes all smaller feature values in the domain and takes them out. The pooling often makes mistakes in obtaining the feature values: first, the variance of the estimate increases; second, the shift of the mean of the estimate. In terms of the prevailing theory, in image processing, the first error handling method mostly uses the mean pooling operation to moderate the size limitation of the domain to reduce the variance, thus making the image background clearer; while the second error handling method mostly uses the maximum pooling operation, which basically ignores the parameter error of the convolutional layer and guarantees the mean accuracy, thus preserving the texture of the image. Therefore, one of these two methods is missing in convolutional neural networks.

**Flatten layer and fully connected layer**

The role of the flatten layer is to flatten multidimensional data into one-dimensional data. The fully-connected layer limits the dimensionality of the data, and thus flattening the data for re-input is essential.

The fully connected layer is often used as the closing layer in the convolutional neural network structure, using different activation functions to match different classification requirements.

**Output Layer**

The role of this layer is to output the final target result.

**Structure of convolutional neural networks [26]**

The layers introduced above are combined to become the complete convolutional neural network structure [27]. Figure 4 shows the basic structure of a CNN, where each convolutional layer applies an activation function for quadratic sampling and then two fully connected layers to give predictions.

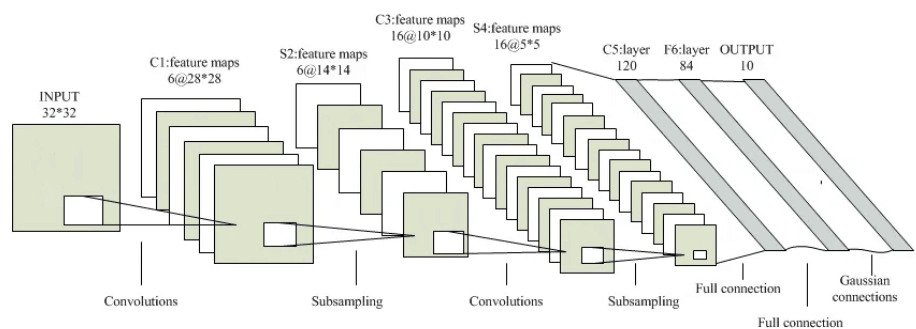

**Fig. 4.** Basic structure of CNN

## 2.3  Common models of convolutional neural networks

Many models of convolutional neural networks exist, and several commonly used models will be presented here.

**LeNet**

LeNet is mainly used to identify and classify non-printed fonts, and it has an accuracy rate of 98%. As a result, the United States put this model into use in the financial industry in the late 20th century. This model is used as the basis of convolutional neural network, with a total of six layers of network, and the convolutional kernels are all 5×5 with a step size of 1,using average pooling: conv → pool → conv → pool → conv(fc) → fc.

**AlexNet**

This model uses the ReLU function as the activation function, and optimizes the problem that the gradient of the sigmoid function is prone to be uncomputable in a network with more layers. And some improvements are made in the final fully connected layer, where only some neurons are randomly selected to participate in the computation of the network, which can prevent overfitting.

Convolutional neural networks usually use average pooling and maximum pooling alternately, but in this model, only maximum pooling is used, basically ignoring the parameter error of the convolutional layer and the size limitation of the neighborhood. This model reduces the step size to achieve a pooling kernel size larger than the step size value, so the output of the pooling layer enhances the feature richness.

A local response normalization layer is created for the first time, so that the neuron responses in this layer show bipolarity and improve generalization ability.

**VGGNet.**

The LRN layer used in AlexNet was not found to bring significant performance improvement to the network in later practice, so the LRN layer in VGGNet has no performance gain (A-LRN) and is not extended to other network models.

VGGNet increases the number of network layers compared with other previous networks, and the number of layers in its network structure is twice or more than AlexNet without counting the pooling and softmax layers here. The concept of convolutional block is proposed for the first time, and 2~3 convolutional layers form a convolutional block, which can reduce the number of parameters and enhance the learning ability by using ReLU activation function.

**GoogLeNet.**

Inception V1 increases the convolution module function compared to several previously proposed network structures. The previous network structure improves the training effect, but the effect benefits from its increased number of network layers also deepens the network depth. However, the deeper depth also brings many problems, such as overfitting, gradient cannot be found in the network and the computational effort increases.

**SqueezeNet.**

SqueezeNet's model compression uses 3 strategies.

(1) replacing 3×3 convolution with 1×1 convolution: the number of parameters of convolution is reduced to 1/9 of the original one, which helps to improve the speed of network operation; (2) reducing the number of channels of 3×3 convolution: the computation of a 3×3 convolution is 3×3×a×b (where a, b are the number of channels of input Feature Map and output Feature Map respectively), reducing the number of channels to reduce the number of parameters The number of channels is reduced to reduce the number of parameters, which helps to simplify the operation and improve the performance of the network; (3) the downsampling is set back: the larger Feature Map contains more information, so the downsampling is moved to the classification layer. Such an operation can improve the accuracy of the network, but it will increase the burden of network computation.

**ResNet.**

Before introducing the model, it is necessary to understand the concept of residuals, first of all, it is necessary to distinguish between residuals and errors. The error is the measured value minus the reference value, and the residual is the difference between the actual observed value and the predicted value, and the residual can detect whether the prediction is accurate or not. The function of one layer in the residual network is set as y=F(x), and the residual model can be expressed as H(x)=G(x) + x, that is, G(x)=H(x)-x. In the unit mapping, y=x is the actual observed value, and H(x) is the fitted value, so G(x) corresponds to the residual, so it is called the residual network.

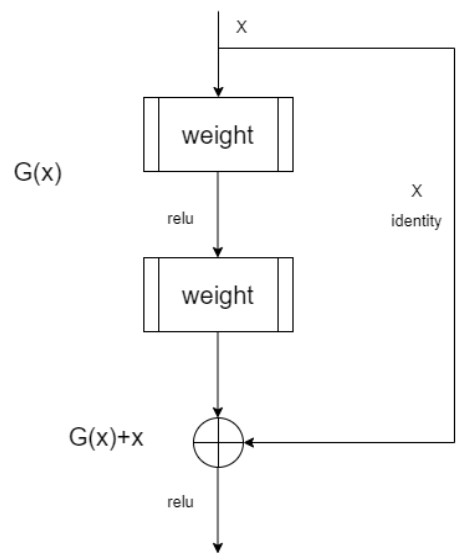

**Fig. 5.** Residual network

Losing the residuals, as shown in the connection on the left side of the figure, the error in training and the network depth show a negative correlation as the number of networks increases. In contrast, theoretically, the increase of network depth and the model training effect should show a positive correlation. Theoretical and practical deviations often exist, and for an ordinary network without jump connections, the deeper the depth will make the computation more complicated, and the improvement and enhancement of the algorithm will be more difficult to achieve. Therefore, in reality, there is a positive correlation between the depth of the network and the number of training errors.

To solve this problem, the network needs to detect the existence of redundant layers by itself, which makes the optimization algorithm complicated and does not achieve constant mapping. The ResNet model is able to solve this problem in a very fitting way by updating the parameters of the redundant layers with the residual $G(x)=0$ instead of the fitted value $H(x)=x$, and by doing so, updating the parameters of the redundant layers. That is, after the network spontaneously detects and infers which layers are redundant and useless, the residual function $G(x)=0$ makes the network of that layer, after removing the redundant layer, match the input of the previous layer accurately. In this way, the effect of errors caused by redundant layers is almost eliminated, effectively solving the network degradation problem.

As an example to explore the cause of network degradation, when one designs the network in the first place, one does not perform the actual operation to grasp the number of layers needed for the network structure. To be on the safe side and to enable the network to train well, people tend to set up more layers of network structure. When the network is actually trained, it is found that only half the number of layers may be needed to complete the task of this network, and then the extra layers are redundant. Therefore, we hope that during the training process, the model can find out that the other half of the layers are redundant and make a constant mapping for only half of the layers, so that the input data will be identical to the output data after passing through the model.

But often the model is likely to learn this half of the constant mapping incorrectly, so it may not work as well as a model with 2/3 of the original number of layers set. Therefore, as the number of layers of the network increases, the effect of model training may degrade, which is caused by the redundant layers learning the wrong constant mappings.

**DenseNet.**

In a comprehensive view, DenseNet has the following advantages over the previous models.

(i) the use of dense connectivity, which mainly improves the back propagation speed of gradients to accelerate the training of convolutional neural networks. (2) The parameters are reduced and the values are decreased to improve the efficiency of computation and to reduce the feature maps specific to each layer; (3) Feature reuse is used to reuse the low-level features for the last layer to play the role of classification.

**MobileNet.**

①MobileNet-v1

In a nutshell, V1 replaces the usual convolutional layers in vgg with depth-separable convolution, and therefore can greatly reduce the number of parameters; and adds the hyperparameters α and β on top of vgg.

②MobileNet-v2

MobileNetV2 is proposed by Google in 2018, with better accuracy and smaller model compared to V1. The model highlights have Inverted Residuals structure (Inverted Residuals) and Linear bottlenecks.

**Deep Residual Learning.**

The core difference of this algorithm is that it proposes a new structure with a topological spreading to form a new block structure, replacing the convolutional block structure of the previous model, which can optimize the performance of the model prediction and improve the accuracy while adding almost no new parameters. The topological spreading also reduces the number of hyperparameters and improves the generality of the model.

**ShuffelNet.**

①ShuffelNet-v1

ShuffleNet is improved by two new operations: point-state group convolution and channel scrubbing, similar to the previous model, which can ensure the accuracy of the network structure output results and reduce the computational complexity. The basic cell structure of the model is optimized and improved based on the residual model cells.

②ShuffelNet-v2

The number of neurons in this model is relatively small, and the number of branches between layers is thus reduced to speed up the model convergence. The model input speed depends on the number of input and output feature channels, but too many grouping parameters can affect the model convergence speed.

**EfficientNet.**

Convolutional neural networks are usually built after resource evaluation, and the more resources are available, the better the performance of the network model will be. This model delves into how to scale the model up and down and finds that making the depth and width of the network converge across the layers or reducing the gap in resolution can both improve the network's effectiveness. Therefore, a new method is proposed to balance the above three characteristics of the network with composite coefficients, etc.

This model was born out of the desire to find a new balance between network depth, width and resolution to measure the accuracy of the network. Previous models have used only one of these aspects to evaluate the effectiveness of the network. This model found that these three aspects together have an impact on the scaling of the network, and explored the evidence of the interaction between the three, based on which the best combination of the three was found.

## 2.4   General Methods for Gradient-Based Attacks

**Membership inference.**

Membership inference [28] refers to speculating whether these data points have been used in the process of training the model based on the known training model and the delimited range of data points. In federation learning, the updated gradient information is fed back to the server every round, so the server is able to have certain local model information. With this attack algorithm, the server is able to know whether the delimited data points are used for model training or not. Sometimes, in certain situations, this attack can directly lead to a privacy breach. For example, if the attack learns that a patient's clinical records are used for training a model for a particular disease, the fact that the patient has that disease is compromised. In practice, Melis et al. demonstrated that this attack approach is extremely accurate on the FourSquare location dataset [29] and can almost determine whether a particular data point data point is used for category classification training.

**Attribute inference.**

Attribute inference refers to inferring whether the corresponding training set contains the same labeled attributes as the known model based on the known training model. Note that the attribute is not important in terms of its relevance to the main task. When training a model on the LFW dataset [30] for identifying gender or race, attribute inference can infer whether they wear a mask or not, in addition to the two known labels. In practice, this also poses a potential risk of privacy compromise. If the patient's age, gender, race, and whether they wear a mask or not are known, there is a high risk that the patient's personal information will be compromised, even if the name and clinical records remain confidential.

**Model inversion.**

Model inversion is a greater threat to the privacy of the training dataset compared to the first two aggressive ones. Since the learning process is always ongoing, this attack exploits this property by having the adversary train a generative adversarial network (GAN) [31] to generate samples that match the training dataset. The results of the attack show that the images obtained are almost identical to the original images, since the GAN is able to create matching samples that are nearly identical to the original training dataset. Moreover, the higher the similarity of the training set, the better the performance of this attack.

The above three attack strategies reveal that the information in the gradient is at risk of leakage to some extent, but each of these three attacks has its own limitations. The membership inference attack relies on delimited data, and the attack will be much more difficult when the input data is not textual information (e.g., images, voice). Attribute inference relaxes the constraint that only a label is needed to perform the attack. However, the attack result will narrow the scope and there is no guarantee to find the specific data. For model inversion, although it can generate synthetic images directly from the statistical distribution of the training data, the results are similar alternatives (rather than the original data) and only work when all class members are similar. What will be investigated and demonstrated in this paper is how to steal the training data completely from the gradient information without prior training data.

## 2.5 Summary of this chapter

This chapter introduced the types of networks and their structures used in this attack. The first section starts with the federal learning system and outlines how it updates the model by gradients; the second section describes the working principle of convolutional neural networks suitable for training classification images and the structure of each level; the third section briefly describes the commonly used convolutional neural network models and provides the basis for the next study on how to select and apply such models for training; the fourth section introduces the The fourth subsection introduces some methods that can be used to perform gradient attacks with prior knowledge of the training data. The theoretical foundation is laid for the subsequent research in this paper to prove the attack algorithm based on gradient features only.

## 3    Design of reconstruction attack algorithm based on gradient features

The subject under study is a reconstruction attack based on gradient features, using a convolutional neural network for the training of a federal learning system for image classification. In this paper, we need to use the gradient derived from the image and its label information trained by the convolutional neural network to restore the original information. This chapter first introduces the principle of the attack that can obtain part of the original data, and then delves into the analysis and study of the algorithm that restores the complete original information based on the gradient.

## 3.1    Local leakage of specific layers

First, this chapter starts with a few special layers to study and optimize the attack algorithm step by step. The first one is the fully-connected layer (FC). The fully connected layer is indispensable in both neural networks and convolutional neural networks. For the biased fully connected layer, it is mathematically proven that the reduction of the original input data from the gradient information is done without considering the position of this layer and the class of layers before and after this layer.

Lemma 1: Suppose a fully connected layer of a neural network contains weights and biases with input $X \in \mathbb{R}^n$ and output $Y \in \mathbb{R}^m$, weight $W \in \mathbb{R}^{m \times n}$ and bias $B \in \mathbb{R}^m$, then it is obtained

$$Y = WX + B \quad (3\text{-}1)$$

If there exists $\frac{dL}{d(B_i)} \neq 0$, then the input data X can B be reconstructed from $\frac{dL}{dW}$ and $\frac{dL}{dB}$. The following proof is carried out: it is known that $\frac{dL}{d(B_i)} = \frac{dL}{dY_i}$ and $\frac{d(Y_i)}{d(W_i)} = X^T$, then

$$\frac{dL}{d(W_i)} = \frac{dL}{d(Y_i)} \cdot \frac{d(Y_i)}{d(W_i)} = \frac{dL}{d(B_i)} \cdot X^T \quad (3\text{-}2)$$

where $Y_i$ $W_i$ and $B_i$ denote the ith row of output *Y*, weight *W* and bias *B*. Therefore, the input X can be reconstructed from this formula as long as $\frac{dL}{d(B_i)} \neq 0$ is satisfied.

The derivative as well as the bias $\frac{dL}{dB}$ are crucial for reconstructing the input layer. To make the gradient attack more general, Geiping et al. delved deeper and found that if the bias *B* is eliminated, the original input data can also be restored from a small amount of gradient information as long as a suitable activation function (e.g., ReLU activation function) is found. The proof process is similar, and the reconstruction of the input data in the fully connected layer still works well.

If the function is not derived, the input data information is still implied in the gradient. For example, in the language classification task, the federal learning system generates corresponding gradients only for the words in the input model, and the attack tells which words and phrases were used for model training in each local data set, respectively. The cross-entropy layer in the classification task, on the other hand, can only generate negative gradients for the data with corrected completion labels. This property gives away the true data labels to some extent.

However, there are many more factors to consider when extending from the fully connected layer (FC) to the more complex convolutional layer (CONV), where the number of features in the convolutional layer and the dimensionality of the input occupation are much larger than the size of the gradient values. A parsing reconstruction method like the one in Lemma 1 will no longer be applicable. Modern convolutional neural networks require a more general attack algorithm.

### 3.2  Complete leakage of the gradient

Zhu et al [33] proposed a new and improved algorithmic method that is able to solve the above problem by using neural networks with the same structure and matching gradients to restore the reconstructed original dataset. Thus it can ensure that the dataset is private and non-interoperable, and the generality and attack capability of this method are broader and more powerful than the methods in the previous subsection, and this technique is called Deep Gradient Leakage algorithm (DLG).

DLG is a reconstruction attack based on gradient features. The attacker receives the gradient update $\nabla W_{t,k}$, $k$ from the other participants $k$ in round $t$, in order to obtain the training set $(x_{t,k}, y_{t,k})$ of participant k from the shared exchange information. Figure 3-1 shows how it works in stealing image information: normal participants input an image from the original private data and derive a prediction by the F-model, then use the difference between the prediction and the labeled value to calculate the gradient, which is returned to the participants to update the model. The algorithm first generates a virtual pixel point image with the same size as the real image, and then initializes a virtual label indicating the probability, such as the cat and dog classification explored in this topic, which sets the label value of 0 for the cat and 1 for the dog. then a softmax layer is generated. the DLG iterates the matching of the image and the label on the intermediate local model to compute the virtual gradient. Note that most FL models share the privacy difference module $F（x,W）$ and the weights $W$ by default.

The loss function is set to be the difference between the true gradient and the virtual gradient, and then the squared number is obtained to ensure that the loss function is positive. The key point of this reconstruction attack is to narrow the gap between the real gradient and the virtual gradient by continuously iterating, and then return to the models of both parties, update their respective parameters, and retrain the attacker's model so that the attacker's gradient value can continuously approximate the real gradient value. When the target loss function is close to zero, the virtual data image will also be infinitely close to the original data image.

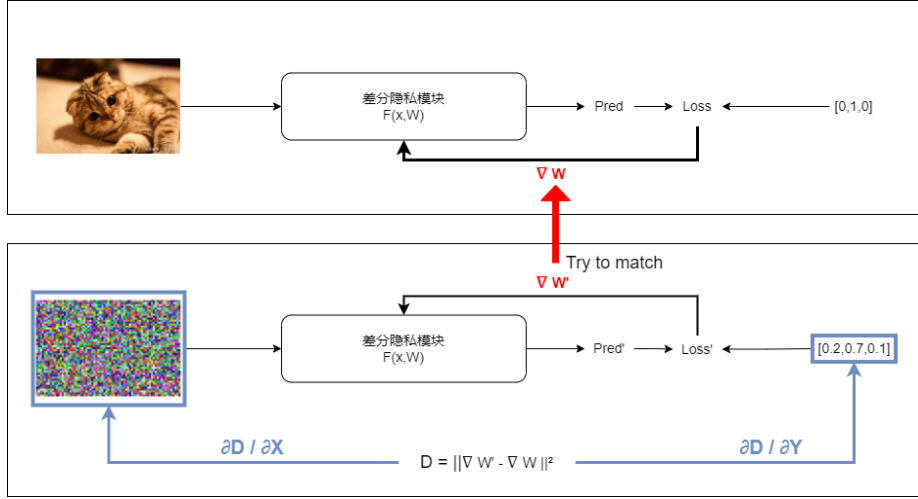

**Fig. 6.** DLG algorithm

In Figure 6, the variables to be updated are marked in bold blue. While the local training model is trained using its differential privacy module and calculates the corresponding $\nabla W$, the attacker uses its own randomly generated input images with label values to derive the gradient $\nabla W'$ and calculates the difference between the two gradients, which the attacker uses as a basis to adjust the parameters and computationally update its virtual input $X$ and label $Y$ so that the gradient loss function converges to a minimum. When the optimization is complete, the attacker can restore the original data information from the local model.

The flow of the algorithm is shown next in mathematical form.

$$\mathbf{x}'^{*}, \mathbf{y}'^{*} = \arg \min_{\mathbf{x}',\mathbf{y}'} \|\nabla W' - \nabla W\|^2 = \arg \min_{\mathbf{x}',\mathbf{y}'} \left\| \frac{\partial \ell(F(\mathbf{x}',W),\mathbf{y}')}{\partial W} - \nabla W \right\|^2 \quad (3\text{-}3)$$

This equation to show how the virtual input $\mathbf{x}'^{*}$ and the label value $\mathbf{y}'^{*}$ 如 are obtained from the gradient reduction.

Let the input be $F\ (\ )$ : the microscopic machine learning model; $W$: the parameter weights; $\nabla W$: the gradient computed from the training data; $\eta$: the learning rate used for DLG optimization. The outputs are the original private training data $x$ and the labels $y$.

① DLG algorithm（$F$，$W$，$\nabla W$）
② $\mathbf{x}'_1 \leftarrow \mathcal{N}(0,1), \mathbf{y}'_1 \leftarrow \mathcal{N}(0,1)$　　　　Initialize virtual inputs and labels.
③ for $i \leftarrow 1$ to $n$ do
④ $\mathbf{L}'_i = \text{softmax}(\mathbf{y}'_i)$
⑤ $\nabla W'_i \leftarrow \partial \ell(F(\mathbf{x}'_i, W), \mathbf{L}'_i) / \partial W_t$　　　　Calculate the virtual gradient.
⑥ $\mathbb{D}_i \leftarrow \|\nabla W'_i - \nabla W\|^2$
⑦ $\mathbf{x}'_{i+1} \leftarrow \mathbf{x}'_i - \eta \nabla_{\mathbf{x}'_i} \mathbb{D}_i$　　　　Update the input data according to the gradient.
⑧ $\mathbf{y}'_{i+1} \leftarrow \mathbf{y}'_i - \eta \nabla_{\mathbf{y}'_i} \mathbb{D}_i$　　　　Update the labels according to the gradient.

15It is important to note that the distance of the gradient, i.e., the loss function $\|\nabla W'_i - \nabla W\|^2$ must be derivable, so that the virtual input data $x$ and label $y$ can be optimized using a standard gradient-based approach. it follows that such optimization requires a second-order derivable function. Here it is assumed that $F$ is a second-order derivable function and this algorithm is applicable to most modern AI models, most neural networks and related tasks.

### 3.3 Optimization of DLG algorithm

The DLG algorithm can restore the complete original data image in most of the scenes, but in this topic, we found that there is a problem that some of the images cannot be restored completely in practice, and we propose an improvement method based on this problem.

Since the original gradient information is generated based on the pixel information of the input image and the label through constant matching, then the richer and more vivid the image color is, the more information the RGB three channels carry, the more pixel information they contain, the more complex the generated gradient is, and the more information can be obtained through the attack, and it is easier to restore the original image. Observe the part of the image that cannot be fully converged, there are mostly large blank areas, which contain relatively less pixel information, so the complete image cannot be restored.

The uneven distribution of image pixel information and the small amount of information in local areas lead to difficulties in image restoration. Thus, the improved algorithm adds the calculation of the average value of the amount of information contained in the image, based on which the hue of the whole image is inferred, and then the variance of each pixel point from the average value is calculated and returned to calculate the gradient and adjust the parameters. When most of the light-colored areas exist in the image, the average value of the image is relatively small, and after other color-rich areas are restored, after iteration, that is, it is possible to calculate the remaining areas based on the average value of the pixel information as light-colored, and to reduce the frequency of random pixel points and dark pixel points to some extent.

### 3.4 Summary of this chapter

Starting from the simplest fully connected layer, this chapter analyzes the principle of reconstructing the input data from the gradient, but this method also has its limitations and is not applicable on CNN networks. Then, an optimization algorithm based on this method is introduced, which not only breaks through the original limitations, but also is better in restoring the original data, and can completely restore the original image and labels based on the gradient. Finally, based on the shortcomings of the DLG algorithm, an improvement method is proposed.

# 4 Performance evaluation of the reconstruction attack algorithm based on gradient features

This chapter shows the implementation of the gradient feature-based reconstruction attack algorithm and the performance evaluation of it and the improved algorithm.

## 4.1 System Environment

The implementation of the attack in this paper is based on the algorithm written in python language, using the self-contained libraries in PyCharm to support the writing of the program, and the libraries, versions, and configurations used are described in Table 1 below.

**Table 1. Software Configuration Description**

| Database | Versions | Description |
| --- | --- | --- |
| opencv-python | 4.5.5.62 | Converts images into pixel information. |
| Pillow | 8.4.0 | Image processing. |
| scikit-learn | 1.0.2 | Contains algorithms such as classification, regression, clustering |
| scipy | 1.7.3 | Differentiation, optimization, image processing |
| tensorboard | 2.8.0 | View training |
| torch | 1.10.1 | Convert data units |
| torchvision | 0.11.2 | Process image data |

The subject is trained on CPU, but the CPU is slow in training images, if conditions allow, it is recommended to use GPU for model training to improve the training efficiency.

This section will compare the DLG algorithm and its improved algorithms, using the two metrics of intuitive image presentation and image restoration as a measure. image restoration This paper uses the mean square error between the restored image and the original image data.

## 4.2 Implementation of reconstruction attacks based on gradient features

**Dogs and cats classification dataset**

The training set of this model uses the cat and dog dataset disclosed by Kaggle in 2013, which consists of 25,000 examples, including 12,500 examples of cats and 12,500 examples of dogs. Therefore, in this paper, 20,000 images are selected as the training dataset and 2,500 as the test dataset. The data consists of RGB three-channel images of various sizes, in which the types of cats and dogs vary in form and the environment they are in, and the label values of cats and dogs are set to 0 and 1, respectively.

**Implementation of DLG algorithm**

The attack process is shown in the figure below. All DLG attacks start with a randomly generated pixel point (the first image) and try to infinitely approximate the generated virtual gradient to the real gradient value. As shown in Table 4-2, the decrease of the mean square error between the virtual image data and the original image data indicates the degree of image convergence, reflecting that the virtual data image gradually approaches the original data image.

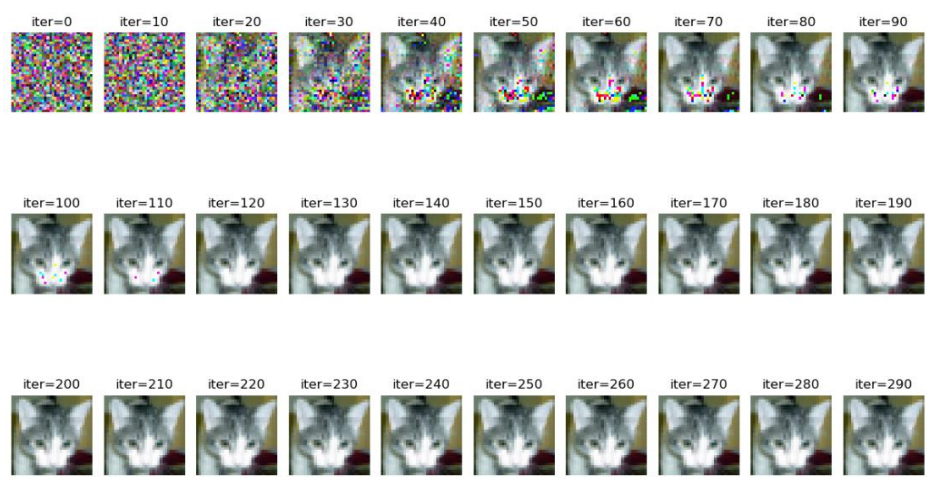

**Fig. 7.** Restore to get the cat picture

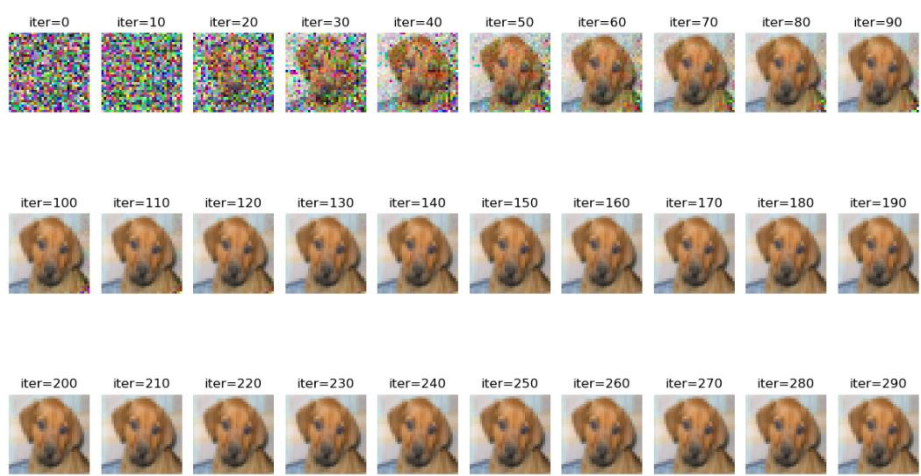

**Fig. 8.** Restore to get the dog picture

**Table 2 Mean square error of the leaked image and the original image**

| Number of iterations | Image | Mean square error |
| --- | --- | --- |
| 20 | 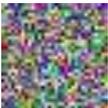 | 105.68 |
| 40 | 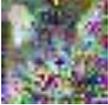 | 99.63 |
| 50 | 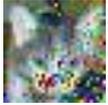 | 89.63 |
| 80 | 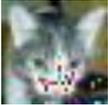 | 54.25 |
| 200 | 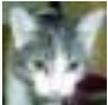 | 3.22 |

## Improved implementation of the algorithm

**Table 3 Comparison of DLG algorithm and improved algorithm**

| Original image | DLG algorithm | DLG Mean Square Error | Improved algorithms | Improved algorithms Mean Square Error |
|---|---|---|---|---|
| 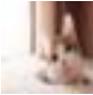 | 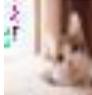 | 24.06 | 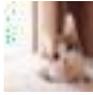 | 19.54 |
| 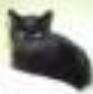 | 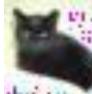 | 47.55 | 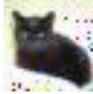 | 42.45 |
| 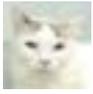 | 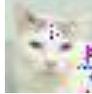 | 40.11 | 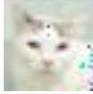 | 25.36 |
| 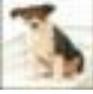 | 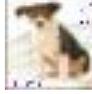 | 28.81 | 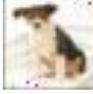 | 24.34 |
| 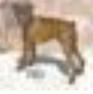 | 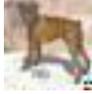 | 30.30 | 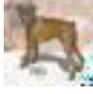 | 22.41 |
| 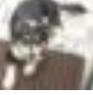 | 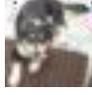 | 28.35 | 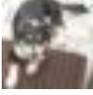 | 15.28 |

From the above table, it can be seen that the number of pixel points present in the images is positively correlated with the mean square error between the images during the restoration of the dog and cat images. It can be visually seen from the image 4rendering effect that the improved algorithm has relatively fewer random pixel points present and the mean squared error between the images and the original image is smaller.

### 4.3 Experimental results and analysis

The DLG attack algorithm used in this paper can attack and restore the vast majority of the original cat and dog pictures based on the gradient, as shown in Figure 7 and Figure 8. Meanwhile, as shown in Table 2, the mean square error between the original data and the original data also tends to the minimum value, which basically stays around 3. However, in the training of a large number of images, it was found that there existed a part of images with poor convergence, which still left randomly generated pixel points. Such images usually have some areas with lighter color nearly white, and after improving the algorithm, as shown in Table 3, it can be observed that the improved algorithm has better restoration of the lighter color areas and the mean square error between the original pixel images is smaller. It illustrates that the reconstruction attack based on gradient features is basically able to restore the local data images in the federal learning system.

### 4.4 Summary of this chapter

This chapter is the implementation and improvement of the gradient feature-based reconstruction attack. The first subsection introduces the programming language used to implement the algorithm, the programming environment, and all the libraries used; the second subsection describes the dataset used and shows the results of the implementation of the attack algorithm in detail; the third subsection analyzes the results and demonstrates that the gradient feature-based reconstruction attack can be a threat to the local data of the federal learning system[34-55].

## 5 Conclusion and Outlook

### 5.1 Conclusion

In this paper, we study the reconstruction attack based on gradient features, mainly using deep learning techniques and algorithms[56-62] for reconstruction attacks. This paper investigates the mechanism of federation learning, the structural hierarchy of convolutional neural networks (CNNs), and the deep gradient leakage (DLG) algorithm that does not rely on the original dataset for the attack.

In this paper, the cat and dog classification dataset is selected as the training model for federation learning, and LeNet, one of the models in CNN, is used for data training. The python language and various libraries in PyCharm are used to complete the reconstruction attack based on gradient features, and the original attack algorithm is improved to make the effect of the restored original image better, which proves that the federation learning gradient has the risk of information leakage.

## 5.2 Deficiencies and problems

In this paper, the gradient-based attack is implemented for the gradient in the federal learning system using relevant techniques, but some problems are found in the implementation and testing sessions of the attack, which need continuous improvement and optimization.

(1) When trying to restore high-resolution images, the attack algorithm is not stable enough, the convergence speed is too slow, and the restoration effect is not good.

(2) When the attack algorithm is applied to images containing only two colors (such as black and white) with large differences, it may fail to converge or converge poorly, and the images have a large number of random pixel points.

(3) The attack algorithm can only do one gradient input to restore an original image for the time being, and cannot input multiple gradients to restore multiple images at the same time.

(4) The current algorithm still has problems such as the applicability is not wide enough, and it cannot attack the training model of text class and so on.

## 5.3 Outlook for follow-up work

Federation learning system will be more widely used in future artificial intelligence technology, although it is not yet seen in some industries, but because of its high efficiency, it must be used more in the future to bring more convenient and fast life to human beings. The research in this paper raises certain questions about the confidentiality of federal learning, and this attack algorithm can be further studied and optimized in depth subsequently.

(1) The DLG algorithm can restore most of the images at present, but there are still some problems, and the follow-up work hopes to continue to improve this algorithm, and improve the convergence speed and accuracy of the restoration of the algorithm.

(2) Different training set categories and training set sizes may affect the training effect and attack effect of the CNN network, which can be supplemented with different categories of images to strengthen the attack algorithm.

(3) This attack algorithm temporarily cannot attack multiple images in batch, and the attack speed is slow, which can be further improved to enhance the efficiency.

## Reference

[1] McMahan, H.B., Moore, E., Ramage, D., Hampson, S., et al. "Communication efficient learning of deep networks from decentralized data", Proceedings of the 20th International Conference on Artificial Intelligence and Statistics, 2017, PMLR 54:1273-1282.

[2] Jochems, A., et al.: Developing and validating a survival prediction model for NSCLC patients through distributed learning across 3 countries. Int. J. Radiat. Oncol. Biol. Phys. 99(2), 344–352 (2017)

[3] Yang, Q., Liu, Y., Chen, T., Tong, Y.: Federated machine learning: concept and applications. ACM Trans. Intell. Syst. Technol. 2019, (TIST) 10(2), 1–19

[4] Krizhevsky, A.: Learning multiple layers of features from tiny images. Citeseer, Technical report ,2009.

[5] Yang Q, Liu Y, Chen T J, et al. Federated Machine Learning[J]. ACM Transactions on Intelligent Systems and Technology, 2019, 10(2): 1-19.

[6] Z Xiao, D Xiao, V Havyarimana, H Jiang, D Liu, D Wang, F Zeng. Toward accurate vehicle state estimation under non-Gaussian noises[J]. IEEE Internet of Things Journal, 2019, 6(6): 10652-10664.

[7] P Zhang, X Chen, X Ma, Y Wu, H Jiang, D Fang, Z Tang, Y Ma. SmartMTra: Robust indoor trajectory tracing using smartphones[J]. IEEE Sensors Journal, 2017, 17(12): 3613-3624.

[8] G. Liu, C. Wang, K. Peng, H. Huang, Y. Li and W. Cheng, "SocInf: Membership Inference Attacks on Social Media Health Data With Machine Learning," in IEEE Transactions on Computational Social Systems, Oct. 2019, vol. 6, no. 5, pp. 907-921.

[9] Pan X, Zhang M, Yan Y, Zhu J, Yang M. Exploring the Security Boundary of Data Reconstruction via Neuron Exclusivity Analysis[J]. arXiv:2010.13356 [cs, stat], 2021.

[10] Zhao Q, Zhao C, Cui S, Jing S, Chen Z. PrivateDL: Privacy-preserving collaborative deep learning against leakage from gradient sharing[J]. International Journal of Intelligent Systems, 2020, 35(8): 1262–1279.

[11] Z. Wang，M. Song，Z. Zhang，Y. Song，Q. Wang and H. Qi, "Beyond Inferring Class Representatives: User-Level Privacy Leak from Federated Learning", IEEE INFOCOM 2019 - IEEE Conference on Computer Communications,2019, pp. 2512-2520.

[12] Jie Li, Fanzi Zeng, Zhu Xiao, Hongbo Jiang, Zhirun Zheng, Wenping Liu, Ju Ren. Drive2friends: Inferring social relationships from individual vehicle mobility data[J]. IEEE Internet of Things Journal, 2020, 7(6): 5116-5127.

[13] L. Melis, C. Song, E. De Cristofaro and V. Shmatikov, "Exploiting Unintended Feature Leakage in Collaborative Learning," 2019 IEEE Symposium on Security and Privacy (SP), 2019, pp. 691-706.

[14] M. Nasr, R. Shokri and A. Houmansadr, "Comprehensive Privacy Analysis of Deep Learning: Passive and Active White-box Inference Attacks against Centralized and Federated Learning," 2019 IEEE Symposium on Security and Privacy (SP), 2019, pp. 739-753.

[15] Fredrikson, M., Jha, S., Ristenpart, T., "Model inversion attacks that exploit confidence information and basic countermeasures", Proceedings of the 22nd ACMSIGSAC Conference on Computer and Communications Security, 2015, pp. 1322–1333.

[16] B. Hitaj, G.Ateniese, Fernando, "Deep Models Under the GAN: Information Leakage from Collaborative Deep Learning", Proceedings of the 2017 ACM SIGSAC Conference on Computer and Communications Security,2017,pp. 603–618.

[17] Salem A, Zhang Y, Humbert M, et al. "Model and Data Independent Membership Inference Attacks and Defenses on Machine Learning Models", 2019 Network and Distributed System Security Symposium, 2019, pp. 1-15.

[18] Jiahui Geng, Yongli Mou, Feifei Li, Qing Li, Oya Beyan. Towards General Deep Leakage in Federated Learning[J/OL]. arXiv:2110.09074[cs], 2021[2022-01-04].

[19] Zhao, B., Mopuri, K.R., Bilen, H.: iDLG: improved deep leakage from gradients. 2020, arXiv preprint arXiv:2001.02610

[20] Geiping, J.; Bauermeister, H.; Dröge, H.; and Moeller, M. "Inverting Gradients - How easy is it to break privacy in federated learning? ", Advances in Neural Information Processing Systems 33,NeurIPS ,2020.

[21] Hongxu Yin, Arun Mallya, Arash Vahdat, Jose M. Alvarez, Jan Kautz, Pavlo Molchanov. "See Through Gradients: Image Batch Recovery via GradInversion", Proceedings of the IEEE/CVF Conference on Computer Vision and Pattern Recognition (CVPR), 2021, pp. 16337-16346

[22] T Liu, JCS Lui, X Ma, H Jiang. Enabling relay-assisted D2D communication for cellular networks: Algorithm and protocols[J]. IEEE Internet of Things Journal, 2018, 5(4): 3136-3150.

[23] D Wang, J Fan, Z Xiao, H Jiang, H Chen, F Zeng, K Li. Stop-and-wait: Discover aggregation effect based on private car trajectory data[J]. IEEE transactions on intelligent transportation systems, 2018, 20(10): 3623-3633.

[24] H Jiang, W Liu, D Wang, C Tian, X Bai, X Liu, Y Wu, W Liu. CASE: Connectivity-based skeleton extraction

in wireless sensor networks[C]//IEEE INFOCOM 2009. IEEE, 2009: 2916-2920.

[25] Ji S, Xu W, Yang M, et al.3D Convolutional Neural Networks for Human Action Recognition[J]. IEEE Transactions on Pattern Analysis & Machine Intelligence,2013,35(1):221-231

[26] Tran D, Bourdev L, Fergus R, et al. "Learning Spatiotemporal Features with 3D Convolutional Network". Proceedings of the IEEE International Conference on Computer Vision (ICCV), 2015, pp. 4489-4497.

[27] X Ma, H Wang, H Li, J Liu, H Jiang. Exploring sharing patterns for video recommendation on YouTube-like social media[J]. Multimedia Systems, 2014, 20(6): 675-691.

[28] Shokri, R., Stronati, M., Song, C., Shmatikov, V."Membership inference attacks against machine learning models",2017 IEEE Symposium on Security and Privacy (SP), 2017, pp. 3–18 IEEE

[29] Yang, D., Zhang, D., Yu, Z., Yu, Z.: Fine-grained preference-aware location search leveraging crowdsourced digital footprints from LBSNs. In: Proceedings of the 2013 ACM International Joint Conference on Pervasive and Ubiquitous Computing, 2013, pp.479–488

[30] Huang, G.B., Ramesh, M., Berg, T., Learned-Miller, E.: Labeled faces in the wild: a database for studying face recognition in unconstrained environments. University of Massachusetts, Amherst, Technical Report 07-49, October 2007

[31] Goodfellow, I., et al.: Generative adversarial nets. In: Advances in Neural Information Processing Systems, 2014, pp. 2672–2680

[32] Geiping, J., Bauermeister, H., Dr¨oge, H., Moeller, M.: Inverting gradients-how easy is it to break privacy in federated learning? 2020, arXiv preprint arXiv:2003.14053

[33] Zhu, L., Liu, Z., Han, S. "Deep leakage from gradients", Annual Conference on Neural Information Processing Systems, (NeurIPS) (2019)

[34] Ping Zhao, Hongbo Jiang, John C. S. Lui, Chen Wang, Fanzi Zeng, Fu Xiao, and Zhetao Li. P3-LOC: A Privacy-Preserving Paradigm-Driven Framework for Indoor Localization. IEEE/ACM Transactions on Networking (ToN), vol. 26, no. 6, pp. 2856-2869, 2018.

[35] Hongbo Jiang, Yuanmeng Wang, Ping Zhao, Zhu Xiao. A Utility-Aware General Framework with Quantifiable Privacy Preservation for Destination Prediction in LBSs. IEEE/ACM Transactions on Networking (ToN), vol. 29, no. 5, pp. 2228-2241, 2021.

[36] Ping Zhao, Jiaxin Sun, and Guanglin Zhang. DAML: Practical Secure Protocol for Data Aggregation based on Machine Learning. ACM Transactions on Sensor Networks, vol. 16, no. 4, pp. 1-18, 2020.

[37] Ping Zhao, Wuwu Liu, Guanglin Zhang, Zongpeng Li, and Lin Wang. Preserving Privacy in WiFi Localization with Plausible Dummy Locations. IEEE Transactions on Vehicular Technology, vol. 69, no. 10, pp. 11909-11925, 2020.

[38] Guanglin Zhang, Anqi Zhang, Ping Zhao, and Jiaxin Sun. Lightweight Privacy-Preserving Scheme in WiFi Fingerprint-Based Indoor Localization. IEEE Systems Journal, vol. 14, no. 3, pp. 4638-4647, 2020.

[39] Ping Zhao, Jie Li, Fanzi Zeng, Fu Xiao, Chen Wang, Hongbo Jiang. ILLIA: Enabling k-Anonymity-based Privacy Preserving against Location Injection Attacks in continuous LBS Query. IEEE Internet of Things Journal, vol. 5, no. 2, pp. 1033–1042, 2018.

[40] Ping Zhao, Hongbo Jiang, Jie Li, Fanzi Zeng, Xiao Zhu, Kun Xie, and Guanglin Zhang. Synthesizing Privacy Preserving Traces: Enhancing Plausibility with Social Networks. IEEE/ACM Transactions on Networking (ToN), vol. 27, no. 6, pp. 2391 – 2404, 2019.

[41] Ping Zhao, Jiawei Tao, Guanglin Zhang. Deep Reinforcement Learning-based Joint Optimization of Delay and Privacy in Multiple-User MEC Systems. IEEE Transactions on Cloud Computing, DOI: 10.1109/TCC.2022.3140231, 2022.

[42] Ping Zhao, Hongbo Jiang, Jie Li, Zhu Xiao, Daibo Liu, Ju Ren, Deke Guo. Garbage in, Garbage out: Poisoning Attacks Disguised with Plausible Mobility in Data Aggregation. IEEE Transactions on Network Science and

Engineering (TNSE), vol. 8, no. 3, pp. 2679-2693, 2021.

[43] Ping Zhao, Xiaohui Zhao, Daiyu Huang, H. Huang. Privacy-Preserving Scheme against Location Data Poisoning Attacks in Mobile-Edge Computing. IEEE Transactions on Computational Social Systems, vol. 7, no. 3, pp. 818-826, 2020.

[44] Ping Zhao, Chen Wang, and Hongbo Jiang. On the Performance of k-Anonymity against Inference Attacks with Background Information. IEEE Internet of Things Journal, vol. 6, no. 1, pp. 808–819, 2018.

[45] Guanglin Zhang, Sifan Ni, and Ping Zhao. Enhancing Privacy Preservation in Speech Data Publishing. IEEE Internet of Things Journal, vol. 7, no. 8, pp. 7357-7367, 2020.

[46] Guanglin Zhang, Anqi Zhang, Ping Zhao. LocMIA: Membership Inference Attacks against Aggregated Location Data. IEEE Internet of Things Journal, vol. 7, no. 12, pp. 11778-11788, 2020.

[47] Guanglin Zhang, Sifan Ni and Ping Zhao. Learning-based Joint Optimization of Energy-Delay and Privacy in Multiple-User Edge-Cloud Collaboration MEC Systems. IEEE Internet of Things Journal, doi: 10.1109/JIOT.2021.3088607, 2021.

[48] P. Zhao, Z. Cao, J. Jiang and F. Gao, "Practical Private Aggregation in Federated Learning Against Inference Attack," in IEEE Internet of Things Journal, 2022, doi: 10.1109/JIOT.2022.3201231.

[49] Hongbo Jiang, Yu Zhang, Zhu Xiao, Ping Zhao and Arun Iyengar. An Empirical Study of Travel Behavior Using Private Car Trajectory Data. IEEE Transactions on Network Science and Engineering, vol. 8, no. 1, pp. 53-64, 2021.

[50] H Jiang, W Liu, D Wang, C Tian, X Bai, X Liu, Y Wu, W Liu. Connectivity-based skeleton extraction in wireless sensor networks[J]. IEEE Transactions on Parallel and Distributed Systems, 2009, 21(5): 710-721.

[51] H Jiang, S Jin, C Wang. Parameter-based data aggregation for statistical information extraction in wireless sensor networks[J]. IEEE Transactions on Vehicular Technology, 2010, 59(8): 3992-4001.

[52] Yennun Huang, Yih-Farn Chen, Rittwik Jana, Hongbo Jiang, Michael Rabinovich, Amy Reibman, Bin Wei, Zhen Xiao. Capacity analysis of MediaGrid: a P2P IPTV platform for fiber to the node (FTTN) networks[J]. IEEE Journal on Selected Areas in Communications, 2007, 25(1): 131-139.

[53] W Liu, D Wang, H Jiang, W Liu, C Wang. Approximate convex decomposition based localization in wireless sensor networks[C]//2012 Proceedings IEEE INFOCOM. IEEE, 2012: 1853-1861.

[54] C Tian, H Jiang, X Liu, X Wang, W Liu, Y Wang. Tri-message: A lightweight time synchronization protocol for high latency and resource-constrained networks[C]//2009 IEEE International Conference on Communications. IEEE, 2009: 1-5.

[55] Y Huang, Z Xiao, D Wang, H Jiang, D Wu. Exploring individual travel patterns across private car trajectory data[J]. IEEE Transactions on Intelligent Transportation Systems, 2019, 21(12): 5036-5050.

[56] K Chen, C Wang, Z Yin, H Jiang, G Tan. Slide: Towards fast and accurate mobile fingerprinting for Wi-Fi indoor positioning systems[J]. IEEE Sensors Journal, 2017, 18(3): 1213-1223.

[57] H Huang, H Yin, G Min, H Jiang, J Zhang, Y Wu. Data-driven information plane in software-defined networking[J]. IEEE Communications Magazine, 2017, 55(6): 218-224.

[58] S Wang, A Vasilakos, H Jiang, X Ma, W Liu, K Peng, B Liu, Y Dong. Energy efficient broadcasting using network coding aware protocol in wireless ad hoc network[C]//2011 IEEE International Conference on Communications (ICC). IEEE, 2011: 1-5.

[59] H Jiang, Z Ge, S Jin, J Wang. Network prefix-level traffic profiling: Characterizing, modeling, and evaluation[J]. Computer Networks, 2010, 54(18): 3327-3340.

[60] H Jiang, A Iyengar, E Nahum, W Segmuller, A Tantawi, CP Wright. Load balancing for SIP server clusters[C]//IEEE INFOCOM 2009. IEEE, 2009: 2286-2294.

[61] H Jiang, P Zhao, C Wang. RobLoP: Towards robust privacy preserving against location dependent attacks in continuous LBS queries[J]. IEEE/ACM Transactions on Networking, 2018, 26(2): 1018-1032.

[62] H Jiang, J Cheng, D Wang, C Wang, G Tan. Continuous multi-dimensional top-k query processing in sensor networks[C]//2011 Proceedings IEEE INFOCOM. IEEE, 2011: 793-801.